\documentclass[fleqn,12pt]{article}
\usepackage{epsfig}
\newcommand{\be}{\begin{equation}}
\newcommand{\ee}{\end{equation}}

\newcommand{\eq}[1]{eq.~(\ref{#1})}

\newcommand{\pbar}{\bar p}

\def\rhonn{\rho_{\rm nn}}
\def\rhogp{\rho_{\gamma\rm p}}
\def\rhogg{\rho_{\gamma\gamma}}

\def\stot{\sigma_{\rm tot}}
\begin{document}    
\renewcommand\thepage{\ }
%
%
\begin{titlepage} 
%
\newcommand\reportnumber{1101} 
\newcommand\mydate{February 14, 2004} 
\newlength{\nulogo} 
\settowidth{\nulogo}{\small\sf{Northwestern-\reportnumber}}
\title{\hfill\fbox{{\parbox{\nulogo}{\small\sf{
Northwestern-\reportnumber\\Brown-HET-1342\\ \mydate %
}}}}\vspace{1in} \\
{A global test of factorization for nucleon-nucleon, $\gamma$p and $\gamma\gamma$ scattering}}
 
\author{
Martin~M.~Block\\
{\small\em Department of Physics and Astronomy,} \vspace{-5pt} \\ 
{\small\em Northwestern University, Evanston, IL 60208}\\
\vspace{-5pt}
\  \\
Kyungsik Kang
\thanks{Work partially supported by Department of Energy contract
DE-FG02-91-Er40688 Task A.}
\vspace{-5pt} \\
{\small\em Department of Physics,}\vspace{-5pt}  \\
{\small\em Brown University, Providence, RI 02912}\\
\vspace{-5pt}\\
}    
\vspace{.5in}
\date{} 
\maketitle
\begin{abstract} 
\noindent The purpose of this note is to show that the cross\ \ section factorization relation $\sigma_{nn}(s)/\sigma_{\gamma p}(s)=\sigma_{\gamma p}(s)/\sigma_{\gamma\gamma}(s)$ is satisfied experimentally in the energy domain $8\le\sqrt s\le 2000$ GeV, where the $\sigma$'s are  total cross sections and $nn$ denotes the even portion of the $pp$ and $\pbar p$ total cross section.   A convenient phenomenological paramaterization for a global {\em simultaneous} fit to the  $pp$, $\pbar p$, $\gamma p$ and $\gamma\gamma$ total cross section data together with the   $\rho$-value data for $pp$ and $\pbar p$ is provided by using real analytic amplitudes.   Within experimental errors, we show that factorization is satisfied when we unfold the published $\gamma\gamma$ data which had averaged the cross sections obtained by using the two different  PHOJET and PYTHIA Monte Carlo results. Our analysis clearly favors the PHOJET results and suggests that the additive quark model, together with vector meson dominance,  allows one to compute $\sigma_{\gamma p}(s)$ and $\sigma_{\gamma\gamma}(s)$ from $\sigma_{nn}(s)$ with essentially no free parameters.  The universal $\rho$-value predicted by our fit, {\em i.e.,} $\rho_{nn}=\rho_{\gamma p}=\rho_{\gamma\gamma}$, is compared to the $\rho$-value obtained by a QCD-inspired analysis of $\pbar p$ and $pp$ data, including the p-air cross sections from cosmic rays.  The $\rho$-values obtained from the two techniques are essentially indistinguishable in the energy region $8\le\sqrt s\le 2000$ GeV, giving us increased confidence in our parameterization of the  cross sections needed for the factorization relation. 
\end{abstract}  
\end{titlepage} 
%
\pagenumbering{arabic}
\renewcommand{\thepage}{-- \arabic{page}\ --}  
%
In this note we investigate experimentally the cross section factorization relation 
\be \frac{\sigma_{nn}(s)}{\sigma_{\gamma p}(s)}=\frac{\sigma_{\gamma p}(s)}{\sigma_{\gamma\gamma}(s)},\label{factorizesigma}
\ee
where the $\sigma$'s are the total cross sections and $\sigma_{nn}$, the total nucleon-nucleon cross section, is the {\em even} (under crossing symmetry) cross section for $pp$ and $\pbar p$ scattering.

Using  eikonals for $\gamma \gamma$, $\gamma p$ and  the even portion of nucleon-nucleon scattering, Block and Kaidalov\cite{bk} have proved the factorization relation of \eq{factorizesigma} by assuming that the ratio of elastic scattering to total scattering is process-independent, {\em i.e.,}
\be
\left(
\frac{\sigma_{\rm elastic}(s)}{\sigma_{\rm tot}(s)}
\right)_{\gamma \gamma}=
\left(\frac{\sigma_{\rm elastic}(s)}{\sigma_{\rm tot}(s)}
\right)_{\gamma p}
=\left(\frac{\sigma_{\rm elastic}(s)}{\sigma_{\rm tot}(s)}\right)_{nn},\quad {\rm for\  all\ }s.\label{sigratios}
\ee
They have further shown that 
\begin{equation} \rhonn(s)=\rhogp(s)=\rhogg(s),\label{eq:rho}
\end{equation}
where $\rho$ is the ratio  of the real to the imaginary portion of the forward scattering amplitude.
These theorems are exact, for {\em all } $s$  (where $\sqrt s$ is the c.m.s. energy), and survive exponentiation of the eikonal (see ref. \cite{bk}). 
The assumption of \eq{sigratios} implies that all of the processes approach a black
disk in the same way.  In the Regge approach, 
factorization breaks down in general, for singularities other than a simple pole 
in the complex angular momentum plane. However, since the radius of interaction for the Pomeron exchange which determines an eikonal grows with energy 
as $R^2 = R_o ^2 +\alpha_P ^{\prime} \ln (s/s_0)$  and the Pomeron slope $\alpha_P 
^{\prime}$ is very small, it was argued by Block and Kaidalov[1] that the 
factorization relations of eq. (1) were expected to be valid to a good accuracy, even in a Regge model.  
Indeed, one can give ``factorization-like'' relations for the residue-like 
constants associated with double or triple poles in the angular momentum plane 
from t-channel unitarity with certain extra provisos [see Cudell {\em et. al.}\cite{cms}].

The strategy of this paper is to test factorization ( \eq{factorizesigma}) empirically  by making a {\em global fit} to all of the experimental data 
for $pp$, $\bar p p$, $\gamma p$, and $\gamma \gamma$ total cross sections and the  $pp$ and $\pbar p$ $\rho$-values, {\em i.e.,} making a simultaneous fit to {\em all} of the available experimental data {\em using} the factorization hypothesis (along with a minimum number of parameters), and seeing if the $\chi^2$ to this global fit gives a satisfactory value.  We find that a convenient phenomenological  framework for doing this numerical calculation is to parameterize the data using real analytic amplitudes that give an asymptotic $\ln^2 s$ rise for the total cross sections\cite{explanation},
for which there are mounting evidences for phenomenological success compared to other forms such as a power form (see for example the COMPETE collaboration\cite{cudell} ). 
We then make the cross sections  satisfy factorization and  test the value of the overall $\chi^2$ to see if the factorization hypothesis is satisfied.    

We will show that the factorization relation $ \frac{\sigma_{nn}(s)}{\sigma_{\gamma p}(s)}=\frac{\sigma_{\gamma p}(s)}{\sigma_{\gamma\gamma}(s)}$ is satisfied experimentally when we use the PHOJET Monte Carlo analysis of the $\gamma\gamma$ cross section data, rather than the published values\cite{L3, OPAL}. We emphasize that the fit using real analytic amplitudes is only phenomenological and is used to provide a convenient analytical structure for the comparison of the {\em shapes} of the total cross sections for the three processes as a function of the energy. 

The COMPETE collaboration\cite{cudell} has also done an analysis of these data, using real analytical amplitudes. However, there are  major differences between our analysis and the one done by the COMPETE group. In order to test factorization, 
\begin{itemize}
\item we fit simultaneously $\pbar p$, $pp$, $\gamma p$ and $\gamma\gamma$ data {\em  assuming complete factorization} using the {\em same} shape parameters, whereas they fit each reaction separately, using {\em different} shape parameters
\item we fit {\em individually} the two $\sigma_{\gamma\gamma}$ sets of L3\cite{L3} and OPAL\cite{OPAL} data that are obtained using the PHOJET and PYTHIA Monte Carlos and do not use their average (the {\em published} value quoted in the Particle Data Group\cite{pdg} compilations), since the two sets taken individually have very different shapes and normalizations compared to their experimental errors. We emphasize that this individual fitting of the $\gamma\gamma$ data, {\em i.e.,} a detailed understanding of the experimental situation, is key to our analysis. 
\end{itemize}
At the end of our computation, we investigate whether the overall $\chi^2$ is satisfactory.

Using real analytic amplitudes, we calculate the total cross sections $\sigma_{nn}$, $\sigma_{\gamma p}$ and $\sigma_{\gamma\gamma}$, along with the corresponding $\rho$-values. This (numerically convenient) technique has a hallowed tradition, being  first proposed by Bourrely and Fischer\cite{bourrely} and utilized extensively by Kang and Nicolescu\cite{kang} and more recently by Block and collaborators\cite{bc,blocketal} 
by Block\cite{rhogp},
by Block and Pancheri\cite{rhogg}, 
and by the COMPETE group\cite{cudell}. This work follows the procedures and conventions used by Block and Cahn\cite{bc}. The variable $s$ is the square of the c.m.s. energy, $p$ is the laboratory momentum and $E$ is the laboratory energy. We will use a scattering amplitude that gives a total cross section that rises asymptotically as $\ln^2(s)$. In terms of the even and odd forward scattering amplitudes  $f^+$ and $f^-$ (even and odd under the interchange of ${\rm E}\rightarrow-{\rm E}$), the even and odd total cross sections $\sigma_{\rm even}$ and $\sigma_{\rm odd}$ are given by the optical theorem as
\be 
\sigma_+=\frac{4\pi}{p}{\rm Im}f^+\quad{\rm and}\quad\sigma_-=\frac{4\pi}{p}{\rm Im}f^-. 
\ee
Thus, 
\be
\sigma_{\pbar p}=\sigma_+ +\sigma_-\quad{\rm and}\quad\sigma_{pp}=\sigma_+ -\sigma_-\label{sigpbarp}
\ee
The cross section $\sigma_{nn}$, referred to in the factorization theorem of \eq{factorizesigma}, is given by
\be
\sigma_{nn}=\frac{4\pi}{p}{\rm Im}f^+,\label{sigmann}
\ee
{\em i.e.,} the {\em even} cross section.
The unpolarized total cross sections for $\gamma p$ and $\gamma\gamma$ scattering are, in turn, given by
\be 
\sigma_{\gamma p}=\frac{4\pi}{p}{\rm Im}f^+_{\gamma p}\quad{\rm and}\quad \sigma_{\gamma \gamma}=\frac{4\pi}{p}{\rm Im}f^-_{\gamma \gamma}\label{sigmagp}.
\ee
In all of the above, the $f$'s and $\sigma$'s are functions of $s$.

We further assume that our amplitudes are real analytic functions with a simple cut structure\cite{bc}. We will work in the high energy region, far above any cuts,  (see ref.\cite{bc}, p. 587, eq. (5.5a), with $a=0$), where the amplitudes simplify considerably and are given by
\begin{equation}
\frac{4\pi}{p}f^+(s)=i\left \{A+\beta[\ln (s/s_0) -i\pi/2]^2+cs^{\mu-1}e^{i\pi(1-\mu)/2}\right\},\label{evenamplitude_nn}
\end{equation}
and
\begin{equation}
\frac{4\pi}{p}f^-(s)=-Ds^{\alpha-1}e^{i\pi(1-\alpha)/2},\label{oddamplitude_nn}
\end{equation}
where $A$, $\beta$, $c$, $s_0$, D, $\mu$ and $\alpha$  are real constants. We ignore any real subtraction constants. In \eq{evenamplitude_nn}, we have assumed that the nucleon-nucleon  cross section rises asymptotically as $\ln^2 s$. 
Using equations (\ref{sigpbarp}), along with \eq{evenamplitude_nn} and \eq{oddamplitude_nn}, the total cross sections $\sigma_{\pbar p}$, $\sigma_{pp}$ and $\sigma_{nn}$ for high energy scattering are given by
\be
\sigma_{\pbar p}(s)= A+\beta\left[\ln^2 s/s_0-\frac{\pi^2}{4}\right]+c\,\sin(\pi\mu/2)s^{\mu-1}-D\cos(\pi\alpha/2)s^{\alpha-1} , \label{sigmapbarp}
\ee 
\be
\sigma_{p p}(s)= A+\beta\left[\ln^2 s/s_0-\frac{\pi^2}{4}\right]+c\,\sin(\pi\mu/2)s^{\mu-1}+D\cos(\pi\alpha/2)s^{\alpha-1} , \label{sigmapp}
\ee 
\be
\sigma_{nn}(s)= A+\beta\left[\ln^2 s/s_0-\frac{\pi^2}{4}\right]+c\,\sin(\pi\mu/2)s^{\mu-1} , \label{sigmann2}
\ee 
and the $\rho$'s, the ratio of the real to the imaginary portions of the forward scattering amplitudes, are given by
\be
\rho_{\pbar p}(s)=\frac{\beta\,\pi\ln s/s_0-c\,\cos(\pi\mu/2)s^{\mu-1}-D\sin(\pi\alpha/2)s^{\alpha-1}}{\sigma_{\pbar p}}\label{rhopbarp},
\ee  
\be
\rho_{pp}(s)=\frac{\beta\,\pi\ln s/s_0-c\,\cos(\pi\mu/2)s^{\mu-1}+D\sin(\pi\alpha/2)s^{\alpha-1}}{\sigma_{pp}}\label{rhopp},
\ee  
\be
\rho_{nn}(s)=\frac{\beta\,\pi\ln s/s_0-c\,\cos(\pi\mu/2)s^{\mu-1}}{\sigma_{nn}}.\label{rhonn}
\ee  
If we assume that the  term in $c$ is a Regge descending term, then $\mu=1/2$.  

To test the factorization theorem of \eq{factorizesigma}, we write the (even) amplitudes $f_{\gamma p}$ and $f_{\gamma\gamma}$ as
\begin{equation}
\frac{4\pi}{p}f_{\gamma p}(s)=iN\left\{A+\beta[\ln (s/s_0) -i\pi/2]^2+cs^{\mu-1}e^{i\pi(1-\mu)/2}\right\},\label{evenamplitude_gp}
\end{equation}
and 
\begin{equation}
\frac{4\pi}{p}f_{\gamma\gamma}(s)=iN^2\left \{A+\beta[\ln (s/s_0) -i\pi/2]^2+cs^{\mu-1}e^{i\pi(1-\mu)/2}\right\}\label{evenamplitude_gg},
\end{equation}
where $N$ is the proportionality constant in the factorization relation $\frac{\sigma_{nn}(s)}{\sigma{\gamma p}(s)}=\frac{\sigma_{\gamma p}(s)}{\sigma{\gamma \gamma}(s)}=N$. 
We note,  from \eq{evenamplitude_nn}, \eq{evenamplitude_gp} and \eq{evenamplitude_gg}, that 
\be
\rho_{nn}=\rho_{\gamma p}=\rho_{\gamma\gamma}=\frac{\beta\,\pi\ln s/s_0-c\,\cos(\pi\mu/2)s^{\mu-1}}{A+\beta\left(\ln^2 s/s_0-\frac{\pi^2}{4}\right)+c\,\sin(\pi\mu/2)s^{\mu-1}}\label{3rhos},
\ee
automatically  satisfying the Block and Kaidalov\cite{bk} relation of \eq{eq:rho}.

In the additive quark model, using vector dominance, the proportionality constant $N=\frac{2}{3}P_{\rm had}^\gamma$, where $P_{\rm had}^\gamma$ is the probability that a photon turns into a vector hadron. Using (see Table XXXV, p.393 of
Ref.~\cite{bauer}) $\frac{f_{\rho}^2}{4\pi}=2.2$,
$\frac{f_{\omega}^2}{4\pi}=23.6$ and $\frac{f_{\phi}^2}{4\pi}=18.4$,
we find
\be 
P_{\rm had}^\gamma\approx\Sigma_{V}\frac{4\pi\alpha}{f_V^2}=1/249,\label{Phadestimate}
\ee 
where $V=\rho,\omega,\phi$. In this estimate, we have neither taken into account the continuum vector channels nor the running of the electromagnetic coupling constant, effects that will tend to increase $P_{\rm had}^\gamma$ by several percent as well as give it a very slow energy dependence, increasing as we go to higher energies.   In the spirit of the additive quark model and vector dominance,  we can  now write, using $N=\frac{2}{3}P_{\rm had}^\gamma$ in \eq{evenamplitude_gp} and \eq{evenamplitude_gg}, 
\be
\sigma_{\gamma p}(s)=\frac{2}{3}P_{\rm had}^\gamma \left(A+\beta\left[\ln^2 s/s_0-\frac{\pi^2}{4}\right]+c\,\sin(\pi\mu/2)s^{\mu-1} \right) \label{sigmagp2}
\ee 
and
\be
\sigma_{\gamma \gamma}(s)=\left(\frac{2}{3}P_{\rm had}^\gamma\right)^2 \left(A+\beta\left[\ln^2 s/s_0-\frac{\pi^2}{4}\right]+c\,\sin(\pi\mu/2)s^{\mu-1} \right) \label{sigmagg2}
\ee  
with the real constants $A,\beta,s_0,c,D$ and $P_{\rm had}^\gamma$   being fitted by experiment (assuming $\alpha=\mu=1/2$).   One might choose to vary the Regge intercepts $\mu$ and $\alpha$ in the fits. Since we want to test the goodness of the cross section factorization relations, we need a reasonable rendition of the {\em even} hadronic amplitude, {\em i.e.}, $\alpha$ plays {\em no} role. Further, a small deviation of $\mu$ from 1/2 also gives no significant difference to our fit to shape of the hadronic data in the energy region of our interest, which is explicitly supported by the results of the COMPETE Collaboration\cite{cudell}.  Total cross sections  for $\gamma p$ scattering have been measured  for energies up to $\approx $ 200 GeV, while  total cross sections  for $\gamma \gamma$ scattering from the OPAL\cite{OPAL} and the L3\cite{L3} collaborations are now available for c.m.s. energies up to $\approx $130 GeV.  

In  fitting the $\gamma\gamma$ data, one might be tempted to use the $\gamma\gamma$ cross sections---along with their quoted errors---that are given in the Particle Data Group\cite{pdg} cross section summary. However, on closer inspection of the original  papers, it turns out that results quoted by the PDG are the {\em averages} of {\em two independent} analyses performed by both the OPAL\cite{OPAL} and L3\cite{L3} groups, using the two different Monte Carlo programs, PHOJET and PYTHIA. The  error quoted by the Particle Data Group was essentially half the  difference between these two very different values, rather than the smaller errors associated with each individual analysis.  

\begin{figure}
\begin{center}
\mbox{\epsfig{file=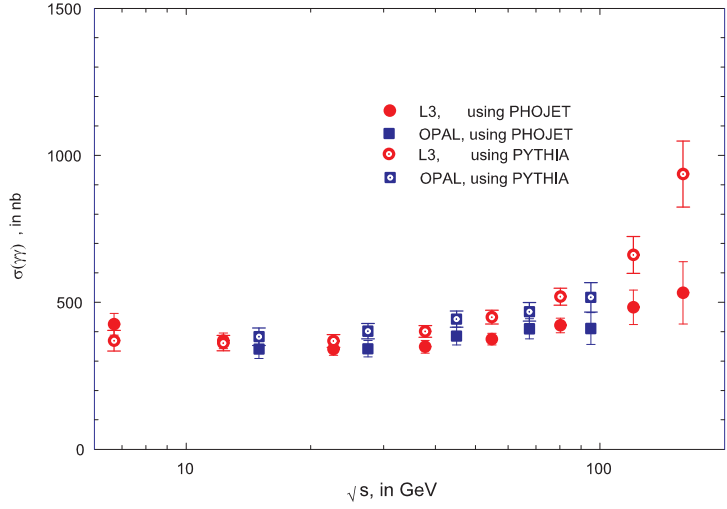,width=4.in,%
bbllx=0pt,bblly=0pt,bburx=438pt,bbury=313pt,clip=}}
\end{center}
\protect\caption[] {\footnotesize OPAL and L3 total cross sections for $\gamma\gamma$ scattering, in nb {\em vs.} $\sqrt s$, the c.m.s. energy, in GeV. The data have been unfolded according to the Monte Carlo used. The solid circles are the L3 data, unfolded using PHOJET and the open circles are the L3 data, unfolded using PYTHIA. The solid squares are the OPAL data, unfolded using PHOJET and the open squares are the OPAL data unfolded using PYTHIA.}
\label{fig:originaldata}
\end{figure}

The Monte Carlo simulations used by OPAL and L3 play a critical role in unfolding the $\gamma\gamma$ cross sections from the raw data.  To quote the OPAL authors\cite{OPAL}, ``In most of the distributions, both Monte Carlo models describe the data equally well and there is no reason for preferring one model over the other for the unfolding of the data.  We therefore average the results of the unfolding.  The difference between this cross section and the results obtained by using PYTHIA or PHOJET alone are taken as the systematic error due to the Monte Carlo model dependence of the unfolding.''  For the testing of factorization, there is good reason for possibly preferring one model over another, since the two models give both {\em different normalizations} and {\em shapes}, which are vital to our analysis. Hence, we have gone back to the original papers\cite{L3, OPAL} and have deconvoluted the data, according to whether PHOJET or PYTHIA was used.  These results are given in Fig. \ref{fig:originaldata}.  Clearly, there are major differences in shape and normalization that are due to the different Monte Carlos, with the PYTHIA results significantly higher and rising much faster for energies above $\approx 15$ GeV.  On the other hand, the OPAL and L3 data agree within errors, for each of the two Monte Carlos, and seem to be quite consistent with each other, as seen in  Fig. \ref{fig:originaldata}. 

For these reasons, we will make three different fits, whose results are shown in Table \ref{ta:amp}.  Fit 1 is a simultaneous $\chi^2$ fit of \eq{sigmapbarp}, (\ref{sigmapp}), (\ref{rhopbarp}), (\ref{rhopp}) and (\ref{sigmagp2}) 
 to the experimental $\sigma_{\pbar p}$, $\sigma_{pp}$, $\rho_{\pbar p}$, $\rho_{pp}$ and $\sigma_{\gamma p}$ data in the c.m.s. energy interval $10 {\ \rm GeV}\le \sqrt s\le 1800$ GeV, {\em i.e.,} we don't include the $\gamma\gamma$ data.   We next make two different simultaneous $\chi^2$ fits of \eq{sigmapbarp}, (\ref{sigmapp}), (\ref{rhopbarp}), (\ref{rhopp}), (\ref{sigmagp2}) and (\ref{sigmagg2}) to the experimental $\sigma_{\pbar p}$, $\sigma_{pp}$, $\rho_{\pbar p}$, $\rho_{pp}$, $\sigma_{\gamma p}$ and the {\em unfolded} $\sigma_{\gamma \gamma}$, using either PHOJET or PYTHIA results, in the c.m.s. energy interval $10 {\ \rm GeV}\le \sqrt s\le 1800$ GeV. Fit 2 uses $\sigma_{\gamma \gamma}$ from PHOJET unfolding and Fit 3 uses $\sigma_{\gamma \gamma}$ from PYTHIA unfolding.  In order to account for possible systematic overall-normalization factors in the experimental data, the  cross sections for L3 are multiplied by the overall-renormalization factor $N_{\rm L3}$ and those for OPAL are multiplied by an overall-renormalization factor $N_{\rm OPAL}$, with these factors  also being fitted in Fits 2 and 3.

>From Fits 1, 2 and 3, we see that the major fit parameters $A$, $\beta$, $s_0$, $D$, $c$ and $P_{\rm had}^\gamma$ are the same, within errors.  The purpose of Fit 1 was to show the robustness of our procedure, independent of the $\gamma\gamma$ data. 

However, when we introduce the unfolded $\gamma\gamma$ cross sections in Fits 2 and 3, we see that the results strongly favor the PHOJET data of Fit 2.  The $\chi^2$/d.f. jumps from 1.49 to 1.87 (the total $\chi^2$ changes from 115.9 to 146.0 for the same number of degrees of freedom). Further, and perhaps more compelling, the normalizations for both OPAL and L3 are in complete agreement, being $0.929\pm0.037$ and $0.929\pm0.025$, respectively.  Thus, they differ from unity by $\approx 7\pm3$\%, compatible with the experimental systematic normalization error of 5\% quoted by L3 . The PYTHIA results from Fit 3 have normalizations that disagree by $\approx 14$\% and $\approx 19$\% for OPAL and L3, respectively, in sharp disagreement with the 5\% estimate. Thus, from here on, we  only utilize the PHOJET results of Fit 2, whose parameters are given in Table 1.
 
Using the parameters of Fit 2, we find that $P_{\rm had}^\gamma=1/(233.1\pm0.63)$, which is in reasonable agreement  with our preliminary estimate of 1/249, being $\approx 6$\% larger, an effect easily accounted for by continuum vector channels in $\gamma p$ reactions that are not accounted for in the estimate of \eq{Phadestimate}.  

\begin{figure}
\begin{center}
\mbox{\epsfig{file=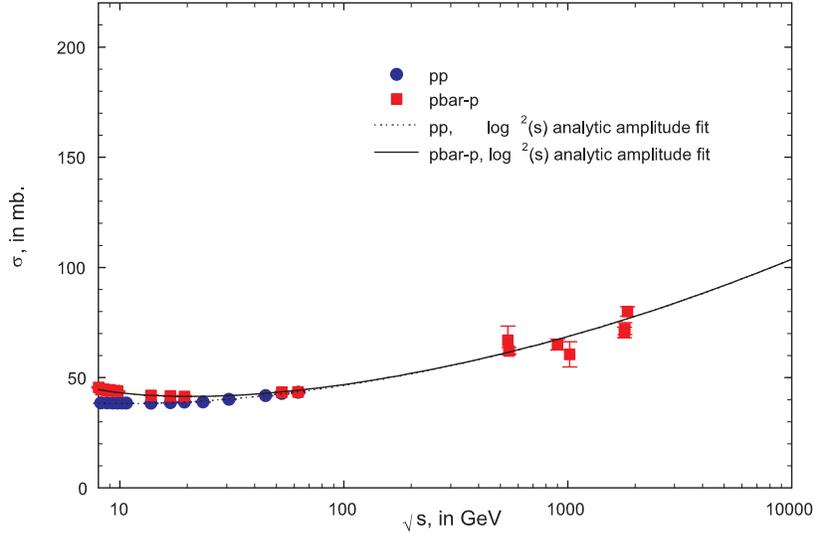,width=4.4in,%
bbllx=0pt,bblly=0pt,bburx=438pt,bbury=313pt,clip=}}
\end{center}
\protect\caption[] {\footnotesize The dotted curve is $\sigma_{pp}$, the predicted total cross section for $pp$ reactions (from Fit 2), in mb,  and the solid curve is $\sigma_{\pbar p}$, the predicted cross section for $\pbar p$ reactions (from Fit 2), in mb {\em vs.} $\sqrt s$, the c.m.s. energy, in GeV. The circles are the experimental data for $pp$ reactions  and the squares are the experimental $\pbar p$ data.
\label{fig:sigmanucleon}}
\end{figure}

\begin{figure}
\begin{center}
\mbox{\epsfig{file=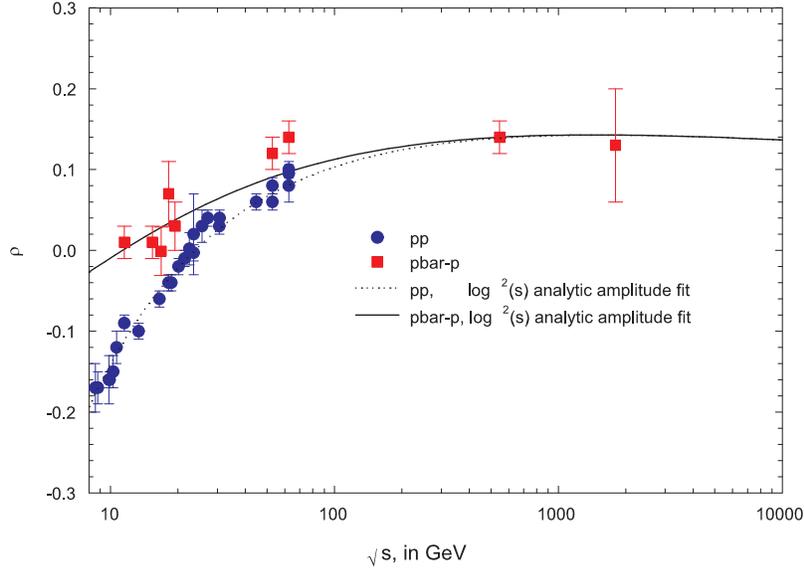,width=4.4in,%
bbllx=0pt,bblly=0pt,bburx=438pt,bbury=313pt,clip=}}
\end{center}
\protect\caption[] {\footnotesize The dotted curve is $\rho_{pp}$, the predicted ratio of the 
real to imaginary part of the forward scattering amplitude for $pp$ reactions and the solid curve is $\rho_{\pbar p}$, the predicted ratio for $\pbar p$ reactions, {\em vs.} $\sqrt s$, the c.m.s. energy, in GeV, from Fit 2. The circles  are the experimental data for $p p$ reactions  and the squares are the experimental $\pbar p$ data.
\label{fig:rhonucleon}}
\end{figure}

\begin{figure}
\begin{center}
\mbox{\epsfig{file=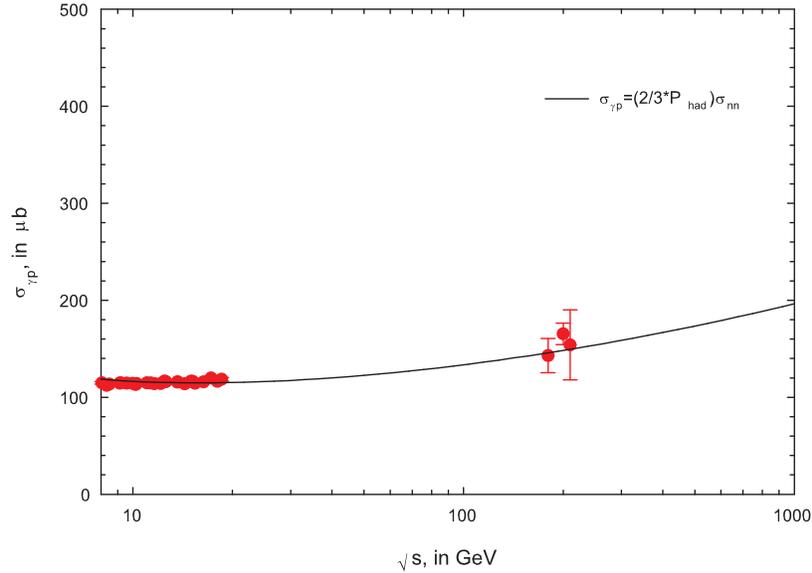,width=4.4in,%
bbllx=0pt,bblly=0pt,bburx=438pt,bbury=313pt,clip=}}
\end{center}
\protect\caption[] {\footnotesize The curve is $\sigma_{\gamma p}=\frac{2}{3}{\rm P}_{\rm had}^\gamma \sigma_{\rm nn}$, the predicted total cross section for $\gamma p$ reactions, in $\mu$b {\em vs.} $\sqrt s$, the c.m.s. energy, in GeV, from Fit 2. The circles are the experimental data. 
\label{fig:sigmagp}}
\end{figure}

The fitted total cross sections $\sigma_{\pbar p}$ and $\sigma_{pp}$ from \eq{sigmapbarp} and \eq{sigmapp} are shown in Fig. \ref{fig:sigmanucleon}, along with the experimental data. The fitted $\rho$-values, $\rho_{\pbar p}$ and $\rho_{pp}$ from \eq{rhopbarp} and \eq{rhopp} are shown in Fig. \ref{fig:rhonucleon}, along with the experimental data. The fitted total cross section $\sigma_{\gamma p}=\frac{2}{3}P_{\rm had}^\gamma\sigma_{nn}$ from \eq{sigmagp2} is compared to the experimental data in Fig. \ref{fig:sigmagp}, using $P_{\rm had}^\gamma=1/233$. The overall agreement of the $\bar p p$, $pp$ and $\gamma p$ data with the fitted curves is quite satisfactory. We now turn our attention to the $\gamma\gamma$ data.

\begin{figure}
\begin{center}
\mbox{\epsfig{file=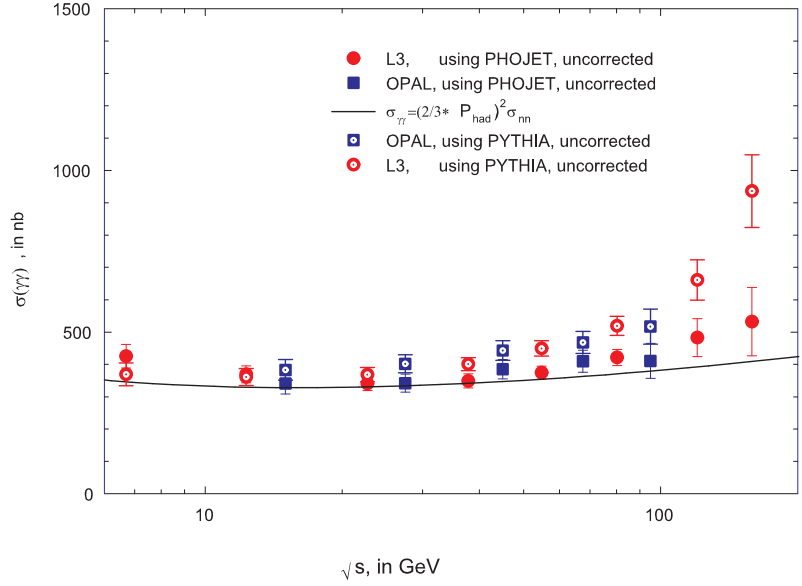,width=4.4in,%
bbllx=0pt,bblly=0pt,bburx=438pt,bbury=313pt,clip=}}
\end{center}
\protect\caption[] {\footnotesize The curve is $\sigma_{\gamma \gamma}=(\frac{2}{3}{\rm P}_{\rm had}^\gamma)^2 \sigma_{\rm nn}$, the predicted total cross section from Fit 2  for $\gamma \gamma$ reactions, in nb, {\em vs.} $\sqrt s$, the c.m.s. energy, in GeV. The open squares and circles are the experimental total cross sections for OPAL and L3, respectively, unfolded using the PYTHIA Monte Carlo. The solid squares and circles are the experimental total cross sections for OPAL and L3, respectively, unfolded using the PHOJET Monte Carlo.
\label{fig:sigmagguncorrected}}
\end{figure}

The fitted total cross section $\sigma_{\gamma \gamma}=(\frac{2}{3}P_{\rm had}^\gamma)^2\sigma_{nn}$ from \eq{sigmagg2} is compared to the experimental data in Fig. \ref{fig:sigmagguncorrected}, again using $P_{\rm had}^\gamma=1/233$. The experimental data plotted in Fig. \ref{fig:sigmagguncorrected} are {\em not} renormalized, but are the results of unfolding the original experimental results, {\em i.e.,} use $N_{\rm OPAL}=N_{\rm L3}=1$. We see from Fig. \ref{fig:sigmagguncorrected} that within errors, both the {\em shape} and {\em normalization} of the PHOJET cross sections from both OPAL and L3 are in reasonable agreement with the factorization theorem of \eq{factorizesigma}, whereas the PYTHIA cross sections are in distinct disagreement. This conclusion is born out by the $\chi^2$'s of Fit 2 and Fit 3 in Table 1.

\begin{figure}
\begin{center}
\mbox{\epsfig{file=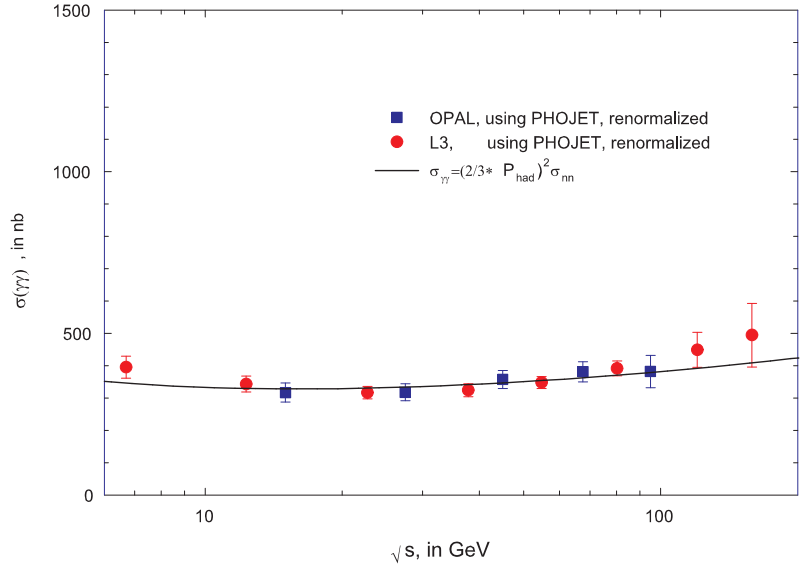,width=4.4in,%
bbllx=0pt,bblly=0pt,bburx=438pt,bbury=313pt,clip=}}
\end{center}
\protect\caption[] {\footnotesize The curve is $\sigma_{\gamma \gamma}=(\frac{2}{3}{\rm P}_{\rm had}^\gamma)^2 \sigma_{\rm nn}$, the predicted total cross section from Fit 2 for $\gamma \gamma$ reactions, in nb, {\em vs.} $\sqrt s$, the c.m.s. energy, in GeV.  The squares and circles are the  total cross sections for OPAL and L3, respectively, unfolded using PHOJET, {\em after} they have been renormalized by the factors $N_{\rm OPAL}=0.929$ and $N_{\rm L3}=0.929$ found in Fit 2 of Table 1.
\label{fig:sigmaggcorrected}}
\end{figure}

Finally, the fitted results for $\sigma_{\gamma\gamma}$, using the parameters of Fit 2, are compared to the renormalized OPAL and L3 (PHOJET only) data in Fig. \ref{fig:sigmaggcorrected}.  The agreement in shape and magnitude is quite satisfactory, indicating strong experimental support  for factorization. 

\begin{figure}
\begin{center}
\mbox{\epsfig{file=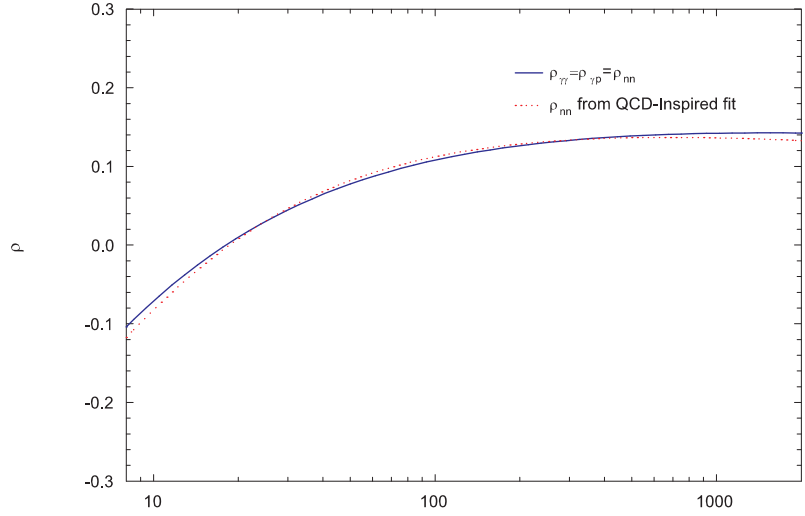,width=4.4in,%
bbllx=0pt,bblly=0pt,bburx=450pt,bbury=319pt,clip=}}
\end{center}
\protect\caption[] {\footnotesize The  solid curve is $\rho_{\gamma\gamma}=\rho_{\gamma p}=\rho_{nn}$, the predicted ratio (from Fit 2) of the 
real to imaginary part of the forward scattering amplitude for the even amplitude {\em vs.} $\sqrt s$, the c.m.s. energy, in GeV. The dotted curve, shown for comparison, is $\rho_{nn}$, the result of a QCD-inspired eikonal fit\cite{blockcr} to $\pbar p$ and $pp$ data that included cosmic ray p-air data.
\label{fig:rho}}
\end{figure}

For completeness, we show in Fig. \ref{fig:rho} the expected $\rho$-value for the even amplitude, from \eq{3rhos}. Also shown in this graph is the predicted value for $\rho_{nn}$ found from a QCD-inspired eikonal fit by Block {\em et al.}\cite{blockcr} to  $\pbar p$ and $pp$ total cross sections and $\rho$-values from accelerators plus p-air cross sections from cosmic rays. The agreement between these two independent analyses, using very different approaches, with one using  
real analytic amplitudes with a $\ln s^2$ behavior and the other using
a QCD-inspired eikonal model in impact parameter space, giving rise to a cross section also eventually rising as $\ln s^2$,
is most striking.  In both cases, these two analyses give  $\rho_{\rm nn}=\rho_{\gamma p}=\rho_{\gamma\gamma}$, another factorization theorem of Block and Kaidalov\cite{bk}.  

We conclude that the cross section factorization hypothesis of \cite{bk}, $\frac{\sigma_{nn}(s)}{\sigma_{\gamma p}(s)}=\frac{\sigma_{\gamma p}(s)}{\sigma_{\gamma\gamma}(s)}$,  is satisfied for nn, $\gamma p$ and $\gamma\gamma$ scattering, if one uses the PHOJET Monte Carlo program to analyze $\sigma_{\gamma\gamma}$. Further, we find that the experimental data also satisfy the additive quark model using vector meson dominance, since 
\begin{eqnarray}
\sigma_{\gamma p}&=&\frac{2}{3}P_{\rm had}^\gamma \sigma_{nn}
\nonumber\\
\sigma_{\gamma \gamma}&=&\left(\frac{2}{3}P_{\rm had}^\gamma\right)^2 \sigma_{nn}\label{vectordominance},
\end{eqnarray}
with $\kappa=2/3$ and $P_{\rm had}^\gamma=1/233$.

The assumption of Block and Kaidalov\cite{bk} in \eq{sigratios} that $\sigma_{\rm elastic}(s)/\sigma_{\rm tot}(s)$ is  process-independent yields another factorization theorem\cite{bk}
\be 
\frac{B_{nn}(s)}{B_{\gamma p}(s)}=\frac{B_{\gamma p}(s)}{B_{\gamma \gamma}(s)}\label{Bfactorize},
\ee
where the $B$'s are the nuclear slopes for elastic scattering (the logarithmic derivatives of the elastic scattering cross sections $d\sigma_{\rm elastic}/dt$, where $t$ is the squared 4-momentum transfer). 
For $\gamma p$ processes, using vector dominance, the $B$'s are the slopes of the `elastic' scattering reactions 
\be
\gamma +p\rightarrow V+p,\label{elasticgp}
\ee
where the vector meson $V$ is either a $\rho$, $\omega$ or $\phi$ meson. Using the additive quark model, \eq{Bfactorize} implies that
\be
B_{\gamma p}(s)=\kappa B_{nn}(s),\quad\ {\rm where\  } \kappa=\frac{2}{3}\label{kappatwothirds}.
\ee
It has been shown by Block, Halzen and Pancheri\cite{bhp} that a $\chi^2$ fit to the available $\gamma p$ data gives
\be
\kappa =0.661\pm0.008, \label{kappavalue}
\ee
in excellent agreement with the 2/3 value predicted by the additive quark model, again justifying the use of 2/3 in our fits. 

We conclude  that if we determine 
$\sigma_{nn}(s)$, $\rho_{nn}$ and $B_{nn}$ from experimental $\pbar p$ and $pp$ data for $\sqrt s\ge 8$ GeV, we can then predict rather accurately $\sigma_{\gamma p}(s)$, $\rho_{\gamma p}$, $B_{\gamma p}$ and $\sigma_{\gamma\gamma}(s)$, $\rho_{\gamma \gamma}$, $B_{\gamma \gamma}$,  in essentially a {\em parameter-free} way, by using factorization and the additive quark model with vector dominance.   Clearly, this conclusion would be greatly strengthened by precision cross section measurements of  both $\gamma p$ and $\gamma \gamma$ reactions at high energies. 

%

%
%
\begin{table}[h,t]                   
%
\def\arraystretch{1.5}            
\begin{tabular}[b]{|l||l|l|l||}
     \cline{2-4}
      \multicolumn{1}{c|}{}
      &\multicolumn{3}{c||}{$\stot \sim \ln^2(s/s_0)$}
      \\
      \hline
      Parameters&\ \ Fit 1: &\ \ \ \ \ \ \ \ Fit 2:  & \ \ \ \ \ \ \ \ Fit 3: \\ 
& no $\sigma_{\gamma\gamma}$&$\sigma_{\gamma\gamma}$ from PHOJET& $\sigma_{\gamma\gamma}$ from PYTHIA\\\hline
     $A$ (mb)&$37.2\pm 0.81$&$37.1\pm$ 0.87&$37.3\pm$ 0.77\\
     $\beta$ (mb)&$0.304\pm 0.023$&$0.302\pm 0.024$&$0.307\pm 0.022$\\
     $s_0$ (${\rm (GeV)}^2$)&$34.3\pm 14$&$32.6\pm 16$&$35.1\pm 14$ \\
     $D$ (mb${\rm (GeV)}^{2(1-\alpha)}$)&$-35.1\pm 0.83$&$-35.1\pm 0.85$
     &$-35.4\pm 0.84$ \\
     $\alpha$&0.5&0.5&0.5 \\
     $c$ (mb${\rm (GeV)}^{2(1-\mu)})$&$55.0\pm 7.5$&$55.9\pm 8.1$
     &$54.6\pm 7.3$ \\
     $\mu$&0.5&0.5&0.5 \\
     $P_{\rm had}^\gamma$&$1/(233.1\pm0.63)$&$1/(233.1\pm0.63)$&$1/(233.0\pm0.63)$ \\
$N_{\rm OPAL}$&\ \ \ \ \ --------&$0.929\pm0.037$&$0.861\pm0.050$\\
$N_{\rm L3}$&\ \ \ \ \ --------&$0.929\pm0.025$&$0.808\pm0.020$\\
     \hline
     $\chi^2/$d.f.&1.62&1.49&1.87\\
     d.f.&68&78&78\\
 total $\chi^2$&110.5&115.9&146.0\\
     \hline
\end{tabular}
     \caption{\protect\small Fit 1 is the result  of a fit to total cross sections and
     $\rho $-values for $\pbar p$ and  $pp$, along with $\sigma_{\gamma p}$. Fit 2 and Fit 3 are the results of fitting total cross sections and
     $\rho $-values for $\pbar p$, $pp$ and $\sigma_{\gamma p}$, as well as including the $\sigma_{\gamma\gamma}$ data from the OPAL and L3 collaborations. Fit 2 uses the results of unfolding  $\sigma_{\gamma\gamma}$ with the PHOJET Monte Carlo, whereas Fit 3 uses the results of unfolding $\sigma_{\gamma\gamma}$ with the PYTHIA Monte Carlo.  The overall-renormalization factors $N_{\rm OPAL}$ and $N_{\rm L3}$ are also fitted in both Fit 2 and Fit 3. The fitted parameters are the ones that have statistical errors indicated.\label{ta:amp}}
\end{table}
\def\arraystretch{1}  

%
%
%
%
%
\newpage
\end{document}